\documentclass[aps,pra,twocolumn,superscriptaddress,10pt]{revtex4-1}
\pdfoutput=1
\usepackage{graphicx}
\usepackage{amsfonts}
\usepackage{amsmath}
\usepackage{amssymb}
\usepackage{epsfig}
\usepackage{dsfont}

\newcommand{\supp}{Appendix}
\newcommand{\V}{\ensuremath{V}}
\newcommand{\psizero}{\ensuremath{0}}
\newcommand{\psiexact}{\ensuremath{\Psi_s}}

\newcommand{\Hint}{\ensuremath{H_{\mathrm{int}}}}

\newcommand{\Htotal}{\ensuremath{H}}
\newcommand{\Hnon}{\ensuremath{H_0}}
\newcommand{\Heit}{\ensuremath{H_{\mathrm{eit}}}}
\newcommand{\Hph}{\ensuremath{H_{\mathrm{ph}}}}
\newcommand{\Q}{\ensuremath{Q}}
\newcommand{\Nexc}{\ensuremath{N_{\mathrm{exc}}}}

\newcommand{\HdiffD}{\ensuremath{h_{\mathrm{dark}}}}
\newcommand{\HtildediffD}{\ensuremath{\tilde{h}_{\mathrm{dark}}}}
\newcommand{\Ueff}{\ensuremath{U^\mathrm{eff}}}
\newcommand{\Uefftwo}{\ensuremath{\bar{U}^\mathrm{eff}}}

\newcommand{\Gac}{\ensuremath{G_0}}
\newcommand{\hph}{\ensuremath{h_{\mathrm{ph}}}}
\newcommand{\hphtilde}{\ensuremath{\tilde{h}_{\mathrm{ph}}}}
\newcommand{\hdark}{\ensuremath{h_{\mathrm{dark}}}}

\newcommand{\kt}[1]{\ensuremath{\left|#1\right\rangle}}
\newcommand{\br}[1]{\ensuremath{\left\langle#1\right|}}
\newcommand{\com}[2]{\ensuremath{\left[#1,#2\right]}}

\newcommand{\qprod}[2]{\ensuremath{\langle#1|#2\rangle}}

\newcommand{\photonmass}{\ensuremath{m_{\mathrm{ph}}}}
\newcommand{\polaritonmass}{\ensuremath{m_D}}
\newcommand{\omegatrans}{\ensuremath{\omega_{\bot}}}
\newcommand{\omegarot}{\ensuremath{\omega_{\mathrm{rot}}}}

\newcommand{\Deltatilde}{\ensuremath{\tilde{\Delta}}}

\newcommand{\chibar}{\ensuremath{\bar\chi}}

\newcommand{\en}{\ensuremath{\epsilon}}

\newcommand{\psiph}{\ensuremath{\psi_a}}
\newcommand{\psie}{\ensuremath{\psi_e}}
\newcommand{\psir}{\ensuremath{\psi_r}}

\newcommand{\phie}{\ensuremath{\phi_e}}
\newcommand{\phir}{\ensuremath{\phi_r}}

\newcommand{\cavAn}{\ensuremath{a}}

\newcommand{\mat}[1]{\ensuremath{\mathbf{{#1}}}}

\newcommand{\coltwo}[2]{\ensuremath{\begin{pmatrix}#1\\#2\end{pmatrix}}}
\newcommand{\colthree}[3]{\ensuremath{\begin{pmatrix}#1\\#2\\#3\end{pmatrix}}}

\newcommand{\vecr}{\ensuremath{\mathbf{r}}}
\newcommand{\vecx}{\ensuremath{\mathbf{x}}}
\newcommand{\vecy}{\ensuremath{\mathbf{y}}}

\newcommand{\vecA}{\ensuremath{\mathbf{A}}}

\newcommand{\vecrho}{\ensuremath{\boldsymbol{\rho}}}

\newcommand{\veckc}{\ensuremath{\mathbf{k_c}}}
\newcommand{\hc}{\ensuremath{\mathrm{h.c.}}}

\newcommand{\uchicago}{Department of Physics and James Franck Institute, University of Chicago, Chicago, IL}
\newcommand{\stuttgart}{Institute for Theoretical Physics III, University of Stuttgart, Germany}

\begin{document}
\title{Quantum Crystals and Laughlin Droplets of Cavity Rydberg Polaritons}

\author{Ariel Sommer}\affiliation{\uchicago}
\author{Hans Peter B\"uchler}\affiliation{\stuttgart}
\author{Jonathan Simon}\affiliation{\uchicago}
\date{\today}

\begin{abstract}
Synthetic quantum materials offer an exciting opportunity to explore quantum many-body physics and novel states of matter under controlled conditions. In particular, they provide an avenue to exchange the short length scales and large energy scales of the solid state for an engineered system with better control over the system Hamiltonian, more accurate state preparation, and higher fidelity state readout. Here we propose a unique platform to study quantum phases of strongly interacting photons. We introduce ideas for controlling the dynamics of individual photons by manipulating the geometry of a multimode optical cavity, and combine them with recently established techniques to mediate strong interactions between photons using Rydberg atoms. We demonstrate that this approach gives rise to crystalline- and fractional quantum Hall- states of light, opening the door to studies of strongly correlated quantum many-body physics in a photonic material.
\end{abstract}

\maketitle

Each of the synthetic quantum material platforms currently under exploration provides an exquisite window into the physics of condensed matter: ultracold atoms in optical lattices~\cite{bloch2008review,bloch2012review} 
benefit from strong interactions, extraordinary coherence, high-fidelity readout~\cite{Greiner2010Mott,scha2012obse,cheu2015quan,Kuhr2015QGM}, and excellent optical control over lattice geometry~\cite{Esslinger2014Haldane}, enabling studies of the Bose- and Fermi- Hubbard models~\cite{bloch2008review,Greiner2010Mott}, along with precision measurements of bulk physics~\cite{yefs2011expl,ku2012sf,bloch2012review}. Trapped ions have been employed for small- to medium- scale studies of quantum magnetism~\cite{Monroe2010Magnetism,Bollinger2012Magnetism}, and classical crystallization~\cite{mitc1998planar,mort2006ions}. Recently, arrays of coupled microwave resonators have been explored as a promising platform for high-fidelity quantum simulation of dynamical equilibrium lattice physics~\cite{houc2012onch,Houck2012LatticeDisorder,kirc2013obse,Jia2013TI,rous2014obse}. 
Reaching a regime of strong, long range interactions, synthetic magnetic fields, and real-time particle injection would open unique routes to probe the interplay of emergent crystallinity, topology, dynamics and dissipative state preparation in a regime inaccessible in other materials. This work describes a route to these objectives in a two-dimensional platform combining multi-mode optical cavities to control single-particle dynamics, with Rydberg slow light polaritons to induce strong interactions.

\begin{figure}[b]
\includegraphics[width=\columnwidth]{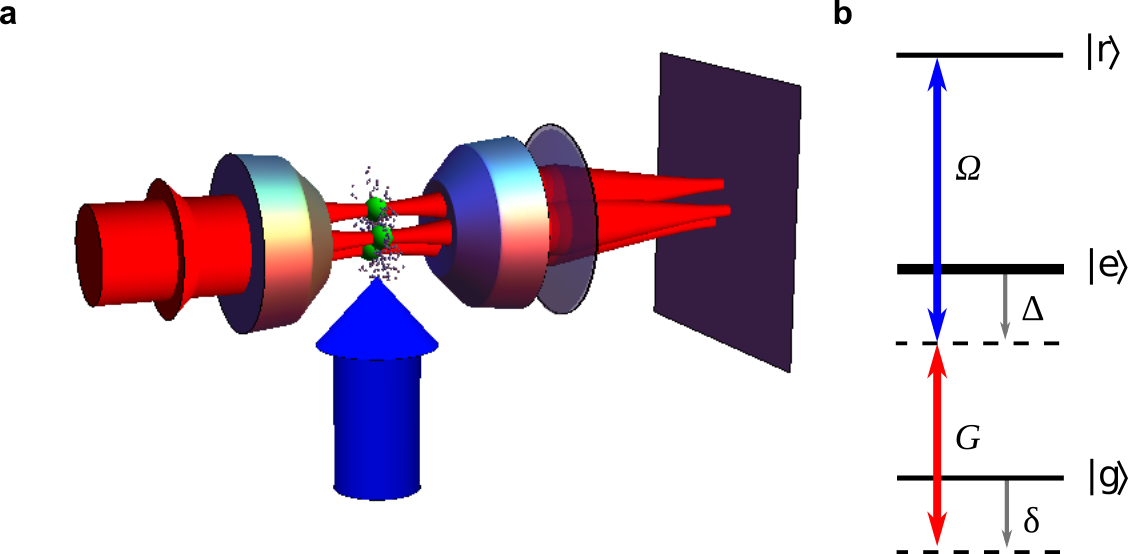}
\caption{{\bf Rydberg polaritons in a multimode optical cavity.} {\bf a}, A quasi-two-dimensional atomic gas of ground-state alkali atoms is held in the waist of an optical cavity formed by a set of mirrors. Photons enter the cavity through a weak driving laser (red) and interact through hybridization with atomic Rydberg excitations, induced by a strong control laser (blue). Information about the state is read out by detecting the transmitted photons. {\bf b}, Relevant atomic level structure giving rise to Rydberg polaritons through electromagnetically induced transparency (EIT). Cavity photons couple the atomic ground state $|g\rangle$ to an intermediate state $|e\rangle$ with collective Rabi frequency $G$, while a strong control laser couples $|e\rangle$ to a Rydberg level $|r\rangle$ with Rabi frequency $\Omega$. The control laser is detuned by $\Delta$ from single photon resonance, and together with the cavity frequency gives a mode-dependent two-photon detuning of $\delta$ from EIT resonance.}%
\label{fig:setup}%
\end{figure}

The combination of electromagnetically induced transparency (EIT) with the strong interaction between Rydberg atoms has recently  emerged as a tool to induce strong interactions between individual photons~\cite{lukin2001dipole,Adams2010rEIT,gors11phot,Pohl2012rEIT,dudi2012rydberg,peyr12quant,stan2013disp,gran14ryd,bien14scattering}.
The appearance of a bound state of photons~\cite{firs2013attract} and the realization of single photon transistors
~\cite{tiar2014sing,gorn2014single} demonstrate the power of this approach, while theoretical proposals have explored the extensions to two-qubit photonic gates~\cite{lukin2001dipole,stan2012gene,otte2013wign,pare2014all}.
In the one-dimensional free-space setups realized so far, the optical depth limits the number of interactions per photon. 
Recently, a single-mode optical cavity was employed to enhance the photon-photon collision probability per photon lifetime, leading to a dispersive nonlinearity~\cite{stan2013disp,gran14ryd}. Meanwhile, theoretical proposals have begun to explore arrays of coupled single-mode cavities to realize photonic lattice physics~\cite{zhan13cavity,magh2015frac}. On the other hand, multimode optical cavities have been proposed as a platform to simulate glassy physics ~\cite{gopa2009emer,kolla2014adj}, and used to observe Bose-Einstein condensation of photons thermalized with dye~\cite{klae2010BEC} or hybridized with excitons~\cite{kasp2006bec,caru13fluids,byrn2014exci}. These systems operate in an open limit where mean-field dynamics of weakly interacting photons can be observed in real-time and particles are injected into the system as desired. It is sensible to consider marrying multimode optical cavities with the strong photon-photon interactions accessible through Rydberg EIT. We explore this possibility, and develop a simple, physical framework for understanding the resulting quantum many-body system.

Our proposed approach employs photons in a family of near-degenerate resonator modes to mimic the physics of a two dimensional gas of massive particles in a trap. The modes must be nearly degenerate so that photons can be coupled between them via Rydberg-mediated interactions, giving rise to photon-photon collisions. In practice, such a setup would consist of a high-finesse optical cavity to engineer the photonic modes, along with a gas of laser-cooled atoms within the cavity to mediate photon-photon interactions (Fig. \ref{fig:setup}). In what follows we provide a formalism that describes the dynamics of the photons as massive trapped particles in the presence of synthetic magnetic fields; introduce coupling to the Rydberg EIT medium, and compute a renormalized photon mass and interparticle potential; and perform numerical experiments demonstrating that few-body phenomena such as crystallization and Laughlin droplet formation are directly observable in such a system.

\section{Cavity Photons as Particles}

\begin{figure}%
\includegraphics[width=\columnwidth]{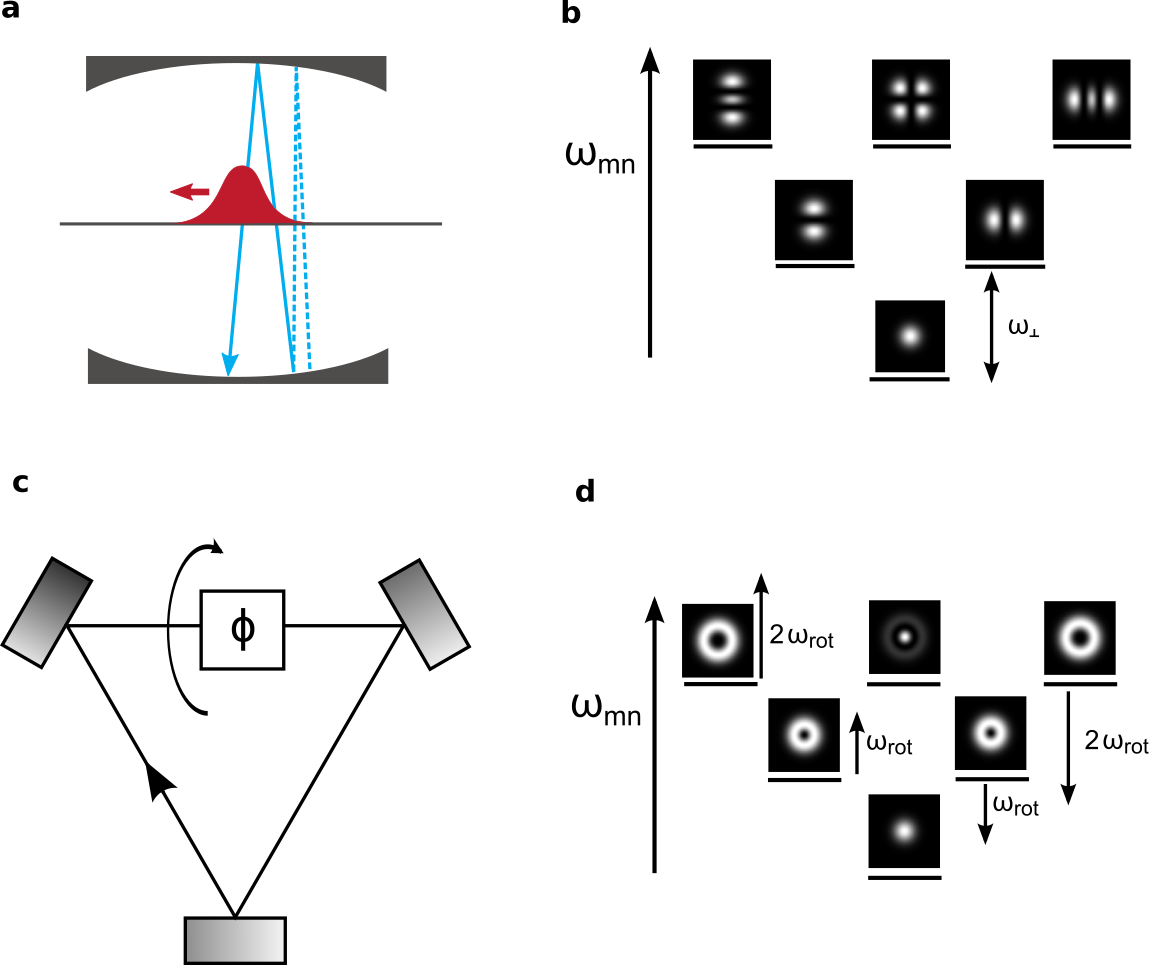}
\caption{{\bf Photons in near-degenerate cavities as particles in two dimensions.} 
{\bf a}, Light rays (blue) intersect the transverse plane of the cavity at slightly different locations after each round trip, owing to the curvature of the mirrors. The intersection points trace out the motion of an effective particle undergoing harmonic oscillation. The cavity eigenmodes in the wave picture correspond to the wavefunction of the particle.
{\bf b}, Mode spectrum of a near-planar cavity within a manifold of transverse modes. The splitting between levels is small compared to the cavity free spectral range (not shown), and approaches zero in the planar limit. 
{\bf c}, Schematic of a helical running-wave cavity that induces an image rotation by an angle $\phi$. 
{\bf d}, Frequency spectrum in a single longitudinal mode of a helical, showing the shift of the mode frequencies proportional to the angular momentum $l$. Modes with negative $l$ are shifted upward in frequency, while modes with positive $l$ are shifted downward towards degeneracy with the TEM$_{00}$ mode, forming a Landau level.  
}%
\label{fig:cavities}%
\end{figure}

It is a remarkable property of nearly degenerate optical cavities that the behaviour of a photon in the transverse plane is well described as a massive particle in an external
potential in two dimensions. This behaviour can be understood in both geometric and wave optics pictures. We begin with geometric optics to provide intuition.

In a degenerate cavity, precise tuning of the geometry causes light rays to retrace the same paths repeatedly, resulting in a fixed set of intersection points with a transverse plane. Tuning slightly away from degeneracy leads to imperfect repetition of the ray paths, causing the intersection points to precess and trace out the path of a particle moving in two dimensions. In the case of a near-planar Fabry-P\'erot cavity with spherical mirrors, where the radius of curvature of the mirrors greatly exceeds the distance between the mirrors, the particle executes harmonic oscillation, as show in Fig. \ref{fig:cavities}a. The transverse oscillation frequency $\omega_\bot$ depends on the cavity geometry, and is independent of the light wavelength. Near degeneracy, $\omega_\bot$ becomes much smaller than the frequency of round trips, allowing one to coarse-grain over the longitudinal motion and consider purely transverse two-dimensional motion.


In the wave optics picture, a photon lives in modes with three quantum numbers: one longitudinal and two transverse. For cavities that are short compared with their mirror radii of curvature, the transverse mode spacing is much smaller than the longitudinal mode spacing, so coarse-graining over the optical round-trip amounts to considering only a fixed longitudinal quantum number. In particular, two-mirror Fabry-P\'erot cavities exhibit Hermite-Gauss (HG) eigenmodes~\cite{sieg1986lasers}, with frequencies $\omega_{mn}=\omega_\bot(m+n+1)+\mathrm{const}$ for the transverse modes HG$_{mn}$, illustrated in Fig. \ref{fig:cavities}b. In the transverse plane of the cavity waist, the HG modes have the same form as the eigenfunctions of the two-dimensional harmonic oscillator, with an oscillator length of $w_0/\sqrt{2}$, where $w_0$ is the cavity waist size ($1/e^2$ intensity radius). This energy- and mode spectrum corresponds to that of the quantum harmonic oscillator, thus the photons in the cavity may be viewed as 2D particles near the quantum ground state of a harmonic trap. The photon ``mass'' then arises from the analogy to zero-point motion, corresponding to the $1/e^2$ intensity radius of the lowest cavity mode $w_0^2=\frac{\lambda}{\pi}\sqrt{L(2R-L)}$, where $L$ is the cavity length, $R$ is the radius of curvature of the cavity mirrors, and $\lambda$ is the wavelength of the light. The photon mass is then $\photonmass = 2\hbar/(w_0^2\omegatrans)$ for HG modes. In the special case of a near-planar cavity, the mass reduces to the relativistic expression $\photonmass=\hbar\omega_{\bot}/c^2$, with $c$ the speed of light~\cite{klae2010BEC}.

More exotic cavity geometries give rise to more complex transverse dynamics of the photons in the focal plane. Of particular interest are geometries that produce an image rotation on each round-trip, arising from a running-wave geometry with either intra-cavity dove prisms, or non-planar geometry (Fig. \ref{fig:cavities}c). Neglecting astigmatism, such helical cavities exhibit Laguerre-Gauss (LG) eigenmodes that carry orbital angular momentum $\hbar l$ in the transverse plane, where $l$ is an integer. The modes are shifted in frequency by $l\omegarot$, with $\omegarot = L_r \phi/c$, where $\phi$  is the round-trip rotation angle and $L_r$ is the round-trip distance. Tuning $\omegarot$ therefore brings modes with different $l$ into degeneracy, as illustrated in Figure \ref{fig:cavities}{\bf d}. The LG modes can be indexed by positive integers $m$ and $n$ counting units of positive and negative angular momentum~\cite{habr07twist} so that $l=m-n$. At degeneracy, the frequency spectrum is independent of one of the indices ($m$). The degenerate manifolds correspond to Landau levels, where the lowest Landau level has $n=0$ and consists of the transverse modes
\begin{equation}
v_{m0}(\vecrho) = \sqrt{\frac{2}{\pi w_0^2 m!}}\left(\frac{\sqrt{2}}{w_0}\right)^m z^me^{-|z|^2/w_0^2}
\label{eq:LLLmode}
\end{equation}
Here $\vecrho=(x,y)$ is the transverse position and $z=x+iy$.  The magnetic length $l_B=w_0/2$ sets the product of the effective charge and effective magnetic field to $4\hbar/w_0^2$, giving 4 flux quanta per mode area ($\pi w_0^2$). The cyclotron frequency $\omega_c$ equals the frequency spacing between Landau levels, determined by the cavity geometry, and determines the mass $\photonmass =4\hbar/(w_0^2 \omega_c)$. Tuning slightly away from degeneracy,
the frequency spectrum becomes $\omega_{mn} = m\omega_\bot^2/\omega_c+n\omega_c$, and induces, in addition to the  magnetic field, a harmonic potential with frequency $\omega_\bot$ that vanishes at degeneracy. The complete coarse-grained Hamiltonian governing the photon dynamics is thus:
\begin{equation}
\hph =  \frac{1}{2\photonmass}\left(-i\hbar\partial_{\vecrho}-\vecA\right)^2 + \frac{1}{2}\photonmass\omega_{\bot}^2\rho^2
\label{eq:HLL}
\end{equation}
with $\vecA=\frac{1}{2}B(-y,x,0)$.

In more general degenerate cavities, light rays only retrace their paths after $s>1$ round trips. Coarse-graining must then also incorporate multiple round trips, resulting in near-degenerate manifolds with mixed longitudinal quantum numbers. The transverse modes in such near-degenerate manifolds span only a subset of the complete Hilbert space, and the resulting photon dynamics exhibit additional symmetries in real- or phase- space. In the case of helical cavities, degeneracies with $s>1$ correspond to particle motion on the surface of a cone~\cite{can2014curved}.
The Methods section provides additional details on the particle description of cavity photons for general degeneracies. 

\section{Coupling to a Rydberg EIT Medium}
Photons do not interact directly with one another, and therefore a photonic material requires a nonlinear medium to mediate photon-photon interactions. Most optical nonlinearities are weak at the few-photon level, making them unsuited to creating strongly-correlated photonic materials. On the other hand, atoms excited to a very large principal quantum number $n\sim 100$ interact very strongly~\cite{saff2010review}. An emerging technology~\cite{gors11phot,peyr12quant} hybridizes photons with Rydberg excitations of ground-state atoms to produce photonic quasiparticles whose interactions come from their Rydberg part, and motion from their photonic part. A second (strong) laser beam helps with this hybridization, as the ground-to-Rydberg oscillator strength is very small, making direct absorption to the Rydberg state very unlikely. The probe photons are absorbed with high probability on a low-lying, nearly closed atomic transition, and from which the atoms are rapidly excited to the Rydberg state by the strong beam. In steady state, these couplings lead to EIT, in which a probe photon propagates as a dark polariton~\cite{flei00dark}. The Rydberg component of the dark polariton then leads to strong polariton-polariton interactions.

We propose using the Rydberg EIT technique described above to hybridize cavity photons with Rydberg excitations of an atomic gas held in the cavity (Fig. \ref{fig:setup}a).
The atomic gas--we consider $^{87}$Rb--is confined to the transverse plane of the cavity waist in a layer that is  thin compared to the Rayleigh range $z_R$ of the cavity waist so that longitudinal diffraction of the photons may be neglected; this will be important to ensure exclusively real-space two-body interactions, as photon diffraction corresponds to a fractional fourier transform~\cite{mend1993frac}. We further require that the sample be sufficiently optically thin as to not mix longitudinal modes of the optical resonator.

A strong control laser couples the first electronic excited state of the atoms $\kt{e}$ to a Rydberg level $\kt{r}$ with Rabi frequency $\Omega$ and detuning $\Delta$, as shown in Fig. \ref{fig:setup}b. Cavity photons are tuned near EIT resonance, allowing them to propagate in the medium as Rydberg polaritons. In the transverse plane, the atomic medium has a uniform density and extends out to a radius that greatly exceeds the cavity waist $w_0$.
Consequently, the atomic medium does not mix different transverse modes of the cavity, except through the induced photon-photon interaction. Individual polaritons therefore have the same qualitative properties as photons in the bare cavity, with rescaled parameters.
The Hamiltonian describing slow light Rydberg polaritons inside a multimode cavity is given by (see \supp{} for details of the derivation):
\begin{equation}
\Htotal=\Hph+\Heit+\Hint
\label{eq:Htotal}
\end{equation}
in which we include non-Hermitian terms to account for dissipation. 
The first term contains the photon dynamics introduced in the previous section,
\begin{equation}
	\Hph = \int d\vecrho\:\psiph^\dag\,\hph\,\psiph
	\label{eq:Hphmain}
\end{equation}
where $\psiph$ is the two-dimensional field operator for photons in a nearly degenerate manifold.
The second term in (\ref{eq:Htotal}) contains the coupling of cavity photons to collective atomic excitations with Rabi frequency $G$, and the coupling to the Rydberg level,
\begin{eqnarray}
	\Heit &=&\frac{\hbar}{2}\int d\vecrho
	\colthree{\psiph}{\psie}{\psir}^\dag\!
	\left(\begin{array}{ccc}
	-i\kappa & G & 0\\
	G  & 2\Deltatilde & \Omega\\
	0 & \Omega & -i\gamma\\
	\end{array}
	\right)\colthree{\psiph}{\psie}{\psir}
\label{eq:Heit}
\end{eqnarray}
where $\psie$ and $\psir$ are the two-dimensional field operators for collective excitations to the atomic excited state and Rydberg state, respectively. Here $\kappa$ is the cavity loss rate, $\Deltatilde=\Delta-i\Gamma/2$ is the complex single-photon detuning, where $\Gamma$ is the atomic excited state decay rate, and $\gamma$ is the Rydberg level decay rate. 
The third term in (\ref{eq:Htotal}) describes the interactions between pairs of Rydberg atoms through a potential $U(\vecr)$,
\begin{equation}
\Hint = \frac{1}{2}\int d\vecr d\vecr' U(\vecr-\vecr')\phi_r^\dagger(\vecr)\phi_r^\dagger(\vecr')\phi_r(\vecr) \phi_r(\vecr')
\label{eq:Hint}
\end{equation}
where $\phi_r(\vecr)$ is the field operator for local excitations of the Rydberg level in three dimensions.
We consider Rydberg S levels giving a van der Waals potential $U(r)=-C_6/r^6$. At high polariton density, additional three-body interactions can appear~\cite{youn2009meso,barr2014ryd}.
Finally, photons enter the cavity through driving by a weak laser at a frequency $\delta_L$ relative to EIT resonance,  which spectroscopically probes the Hamiltonian (\ref{eq:Htotal}) and drives it to a dynamical steady state.

\begin{figure}%
\includegraphics[width=\columnwidth]{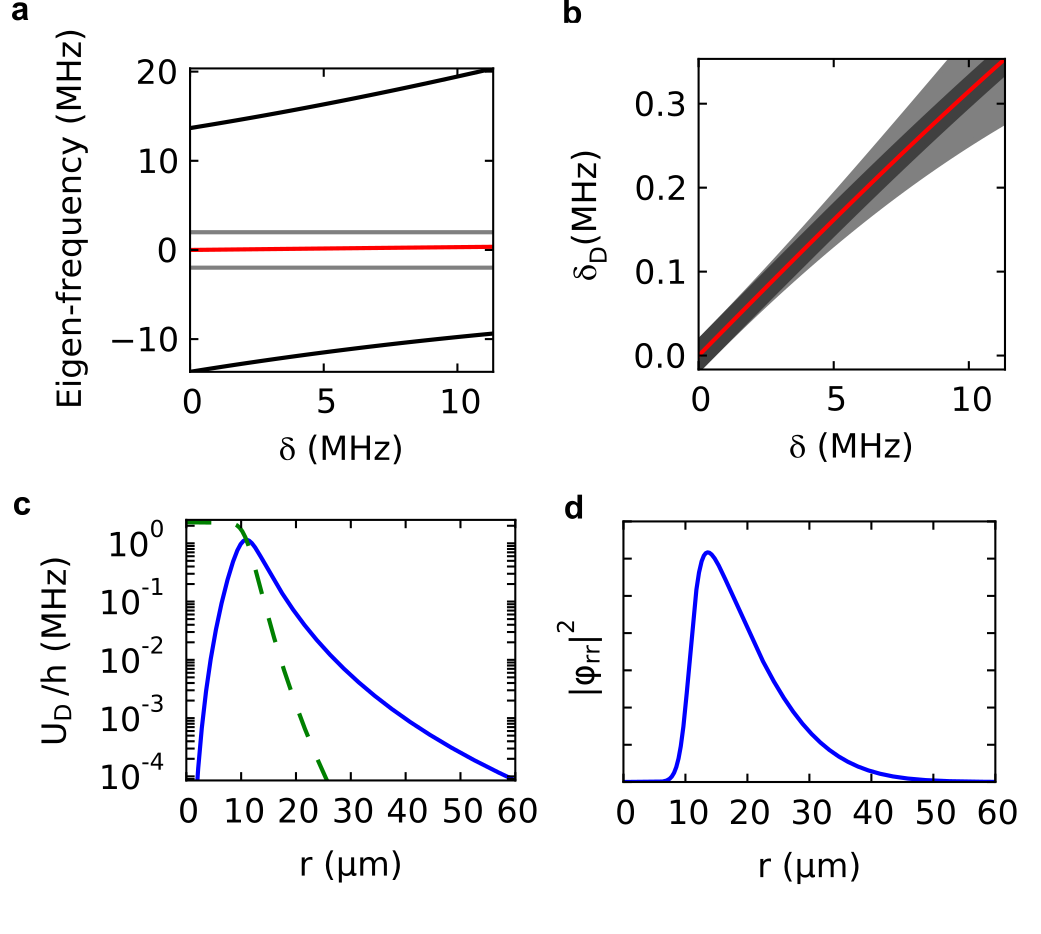}%
\caption{{\bf Dispersion and interaction of cavity Rydberg polaritons.} {\bf a}, Eigenfrequencies of the single-polariton states coupled to a cavity mode with detuning $\delta$ from EIT resonance. Red: dark state, black: bright states, gray: matter-only states, independent of $\delta$, in spatial modes not coupled to the cavity.
{\bf b}, Magnified view of the dark polariton frequency $\delta_{D}$.
The lightly shaded region shows the dark polariton linewidth (imaginary part of the frequency). The dark shaded region shows the contribution of the photon out-coupling.
{\bf c}, Effective potential $U_D$ between two dark polaritons versus distance $r$. The short-range limit is given by the parameter $-1/\chibar$ defined in the text, while the long-range behaviour is proportional to $1/r^6$ for van der Waals interactions. Solid: real part, dashed: imaginary part. {\bf d},
 Rydberg-Rydberg component $\varphi_{rr}$ of the wavefunction for two dark polaritons in a single-mode cavity. The extent of the wavefunction is determined by the cavity waist $w_0$, while the wavefunction is suppressed at short range $r<\xi$, with $\xi$ the blockade radius.  
The parameters are $G=2\pi\cdot 27$ MHz, 
$\kappa=2\pi\cdot 1.0$ MHz, $\Omega = 2\pi\cdot 5$ MHz,
$n=10^{11}\,\mathrm{cm}^{-3}$, 
$w_0=30\,\mu$m, 
$L_r=6.8$ cm, and
$C_6=-2\pi\cdot 4243$ GHz $\mu\mathrm{m}^6$ for the 80S Rydberg level of $^{87}$Rb.}
\label{fig:polaritons}%
\end{figure}

The low-energy excitations of the system consist of dark polariton quasiparticles~\cite{flei00dark,flei05eit}.
A dark polariton has probability $\cos^2\theta$ to be a photon and a probability $\sin^2\theta$ to be a Rydberg excitation, where $\theta=\tan^{-1}(G/\Omega)$ gives the dark state rotation angle. The dark-polariton-projected single-particle Hamiltonian is given by a rescaling of the photon Hamiltonian (see \supp{} for details of the derivation),
\begin{eqnarray}
	\hdark &=& \hph\:\cos^2\theta\nonumber\\
	 &=& \frac{1}{2 m_D}\left(-i\hbar\partial_{\vecrho}-\vecA\right)^2 + \frac{1}{2}m_D\omega_{\bot D}^2\rho^2
	\label{eq:hdark}
\end{eqnarray}
where the polariton oscillation frequency is $\omega_{\bot D}=\omega_\bot\cos^2\theta$ and the mass is $m_D=\photonmass/\cos^2\theta$. 

The Hamiltonian (\ref{eq:hdark}) has discrete eigenstates that each correspond to a dark polariton in a particular cavity mode. The eigenfrequency of a dark polariton, measured relative to the EIT resonance frequency for a probe photon, is then given by $\delta \cos^2\theta$, where $\delta$ is the detuning of the cavity mode from EIT resonance. Likewise, the loss rate of a dark polariton state is given to leading order by $\kappa\cos^2\theta + \gamma\sin^2\theta$, corresponding to out-coupling through the cavity mirrors and loss through decay of the Rydberg state. For larger detunings, breaking of the EIT condition leads to additional loss proportional to $\delta^2$. Figure \ref{fig:polaritons}a-b shows the frequencies of the single-polariton eigenstates of the system, including the dark polariton state as well as four higher-frequency states.

\begin{figure*}[t]%
\includegraphics{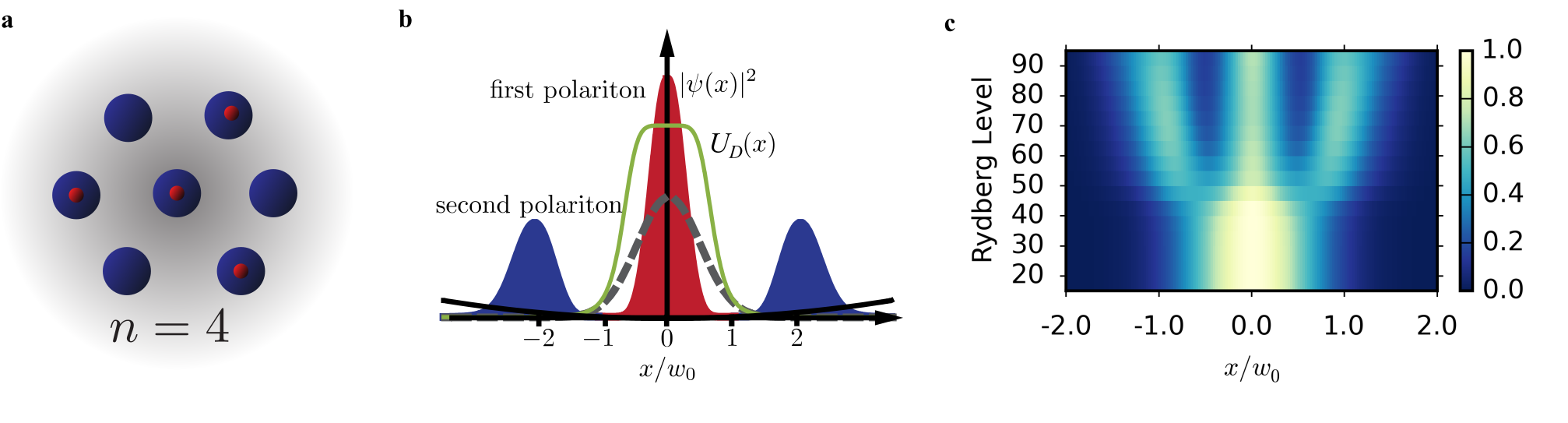}%
\caption{{\bf Few-body emergence of crystalline structures.} The restriction to even harmonic oscillator wavefunctions for a cavity with $s=2$ leads to unique crystals where individual photons occupy a superposition of two lattice sites. {\bf a}, Illustration of the photon probability for four photons in a cavity with $s=2$. The red dots mark possible positions of the photons in the cavity in a single shot. {\bf b} and {\bf c} show numerical calculations for two photons in a cavity with one transverse dimension, implemented using TEM$_{m0}$ modes. {\bf b}, Probability to find a photon versus transverse position in the cavity. One photon (red) is predominantly sitting in the center of the trap, while the second photon  (blue) can be found with equal probability either to the left or to the right. The dashed line illustrates the lowest harmonic oscillator wavefunction, while the green line indicates the effective interaction potential. {\bf c}, Fluid-to-crystal transition of two photons with increasing interaction strength. The color scale shows the emitted light intensity distribution in arbitrary units.}%
\label{fig:crystallinestructure}%
\end{figure*}

The Rydberg-Rydberg interactions detune the Rydberg level at short range, leading to a screening of the divergence of $U(r)$ at $r=0$. 
The screening results from mixing of the higher-frequency single-particle states with the dark state in the two-body wavefunction.
At low energies, the screened potential becomes $\Ueff(r) = -C_6/(C_6\chibar+r^6)$, as in Ref.~\cite{bien14scattering},
where $\chibar$ is given by
\begin{equation}
\hbar\chibar=\frac{\Deltatilde}{2(\Omega/2)^2}-\frac{1}{2\Deltatilde}
\label{eq:chibarzero}
\end{equation}
and provides a characteristic length scale $\xi = |C_6\chibar|^{1/6}$
for the blockade of two Rydberg excitations.
The interaction potential $U_D(\rho)$ between dark polaritons is then determined by restricting to two dimensions and projecting onto the dark polariton basis, as we show in the \supp. 
In the limit where the atomic medium is thin compared to the blockade radius, it reduces to $U_D\approx\alpha \Ueff(\rho)$,
where $\alpha=\sin^4\theta$ is the probability for two dark polaritons to both be Rydberg excitations.
%
Figure \ref{fig:polaritons}c shows the interaction potential for a realistic set of parameters, while Fig. \ref{fig:polaritons}d shows the two-Rydberg part of the wavefunction in a two-polariton state.
%

The length and time scales governing the two-dimensional gas of Rydberg polaritions allow direct measurements on the system using optical techniques. Conventional optics can resolve the blockade radius of typically 10$\mu$m, which sets the smallest distance between polaritons in a stable configuration. Likewise, existing electronics can resolve the timescale set by the EIT linewidth of typically a few MHz.

\section{Crystallization}

While the photonic BEC regime has been explored previously for non-interacting photons thermalized via a dye \cite{klae2010BEC} and in exciton-polariton condensates~\cite{kasp2006bec}, many-photon systems with strong interactions are wholly unexplored. 
Unlike cold atoms, for which interactions overwhelm kinetic energy only in the presence of an optical lattice to increase the effective mass of the atoms, the ability to tune the photon “mass” via the degeneracy of the resonator means that for realistic parameters, the Rydberg interactions can overwhelm the kinetic dynamics of the polaritons, allowing for studies of emergent crystallinity. In a near-planar cavity, the mass and harmonic oscillator frequency have a fixed relation to each other. However, for more general near-degenerate cavities, it is possible to design the harmonic oscillator frequency and the mass independently through the choice of cavity waist.

Crystallization is expected beyond a critical value of the ratio $r_d$ between the interaction energy and kinetic energy in the system~\cite{buch2007polar}. To obtain conditions for crystallization, we assume that the separation $d$ between dark polaritons exceeds the blockade radius $\xi$. Then the effective interaction simplifies to $U_D\approx -\alpha C_6/\rho^6$ and $r_d=\alpha|C_6| \polaritonmass/(\hbar^2 d^4)$. Crystallization therefore occurs at high density, as in the case of dipolar interactions. The density depends on the strength of the driving laser and the harmonic confinement. For self-consistency, and to prevent losses due to detuning from the EIT condition, the critical inter-particle spacing must exceed $\xi$,
\begin{equation}
1 \lesssim \left(\frac{d}{\xi}\right)^{12} = \left(\frac{\alpha \polaritonmass}{r_d \hbar}\right)^3
		\frac{C_6/\hbar}{|\hbar\chibar|^2}
\label{eq:crystalcondition}
\end{equation}
The relation (\ref{eq:crystalcondition}) gives a condition on the polariton mass and therefore on the degree of cavity degeneracy $\omega_\bot$ for a given cavity waist.
%
To obtain concrete values, we need to know $r_d$ at the critical point. Although this quantity is currently unknown for van der Waals interactions, quantum Monte Carlo simulations~\cite{buch2007polar} of crystallization in two dimensions with dipolar interactions give a critical value of $r_d=18\pm 4$; we use $r_d=20$ as an estimate here. For the parameters given in Fig. \ref{fig:polaritons} using the 80S Rydberg level and $\Delta=0$, the condition (\ref{eq:crystalcondition}) then gives $\omega_{\bot} \lesssim 5.8$ MHz, which current technology for optical cavities can readily achieve.

The emergent few-body structures for a nearly degenerate cavity with $s=2$, allowing only even harmonic oscillator wavefunctions, are shown in Fig. \ref{fig:crystallinestructure}. The Hilbert space restriction for $s=2$ results in a unique type of crystalline order, as illustrated in Fig. \ref{fig:crystallinestructure}a. The symmetry of the photonic modes implies that photons localized away from $\vecrho=0$ occupy superpositions of two locations, and these superpositions must fit in with the crystal structure. As a basic check of this idea, we numerically simulate the driven cavity system for up to two photons, and find that when the interaction strength is sufficient, the photons become localized to different positions, indicating the two-photon analogue of crystallization. The three peaks indicated in Fig. \ref{fig:crystallinestructure}b-c result from one photon localizing at $\vecrho=0$ with the other localizing at two symmetric positions. Figure \ref{fig:crystallinestructure}b shows the ground state of a two-polariton system while Fig. \ref{fig:crystallinestructure}c shows the normalized two-photon part of the steady-state wavefunction for the driven system. The calculation in Fig. \ref{fig:crystallinestructure}c uses the same parameters as given in the caption to Fig. \ref{fig:polaritons} except that in 4c, $\Omega = 2\pi\cdot 2.5$ MHz and $\omega_{\bot}=2\pi\cdot 500$ kHz. For clarity, Fig. \ref{fig:crystallinestructure}b uses a larger interaction strength ($\xi=0.7 w_0$) and weaker harmonic trapping $\omega_{\bot}=2\pi\cdot 5$ kHz to increase the separation between photons.

\section{Laughlin Droplets}
\begin{figure}%
\includegraphics[width=\columnwidth]{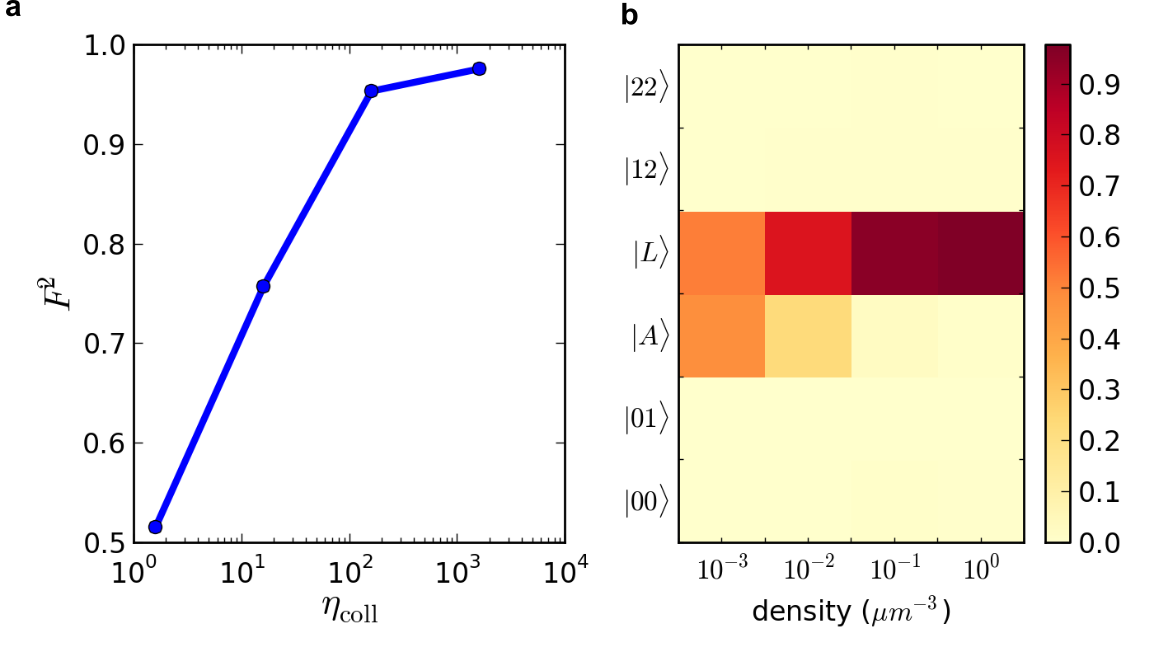}%
\caption{{\bf Laughlin states of light.} {\bf a}, Probability $F^2$ for two emitted photons to occur in the two-boson Laughlin state, versus the collective cooperativity $\eta_{\mathrm{coll}}\equiv G^2/(\kappa \Gamma)$, with photons injected into the $l=1$ mode. {\bf b}, Decomposition of the two-photon wavefunction into the basis $|l_1 l_2\rangle$ with $l_{1,2}$ the angular momentum indices of the two photons. The states $|11\rangle$ and $|02\rangle$ are replaced with the states $|L\rangle,|A\rangle=\frac{1}{\sqrt{2}}(|11\rangle\mp|02\rangle)$, giving the Laughlin (-) and anti-Laughlin (+) states, respectively. The color scale gives the probability for two photons to be in the given state.}%
\label{fig:fqh}%
\end{figure}

For a cavity with Landau-level-like degenerate manifolds, interactions between dark polaritons lead to bosonic Laughlin states~\cite{hafe2013fqh,umuc14probing,magh2015frac} as many-body eigenmodes of the system. 
Unlike solid-state realizations~\cite{Stormer1999FQHE} and cold-atom proposals~\cite{GemelkeFQH,sore2005frac}, where the Laughlin state is realized through thermal phase transitions and adiabatic state transformations, respectively, the openness of the photonic system enables spectroscopically-resolved preparation of the photonic Laughlin state.

Laughlin states arise in the limit of a small blockade radius $\xi \ll w_0$ where the effective interaction reduces to a contact interaction. Additionally, we consider the limit of small mode splitting $\omega_{\bot}\rightarrow 0$ to reach a Landau level. 
When photons are injected into the lowest Landau level with angular momentum $l=0$, strong blockade allows only a single polariton into the system, owing to the repulsion between the intracavity polaritons. When photons are injected into the $l=1$ mode of the lowest Landau level, the blockade is lifted for two photons, as they are able to enter the system in a bosonic Laughlin state~\cite{umuc14probing},
\begin{equation}
\kt{\Psi_{2L}} =\frac{1}{\sqrt{2}}\left[\frac{1}{\sqrt{2}}\left(A_{1D}^d\right)^2 -  A_{0D}^d A_{2D}^d\right]\kt{0}
\label{eq:psiLaughlin}
\end{equation}
with $A_{mD}^d$ the creation operator for a dark polariton in mode $m$. In position space, (\ref{eq:psiLaughlin}) has the
%
spatial wavefunction
\begin{equation}
\qprod{\vecrho_1,\vecrho_2}{\Psi}\propto\left(z_1-z_2\right)^{1/\nu} e^{-(|z_1|^2+|z_2|^2)/w_0^2}
\label{eq:spaceLaughlin}
\end{equation}
with $\nu=1/2$.
The Laughlin state has vanishing contact interaction energy, and non-zero overlap with two photons in the $l=1$ mode. 
Figure \ref{fig:fqh} shows results from a simulation of the driven system in the two-photon Laughlin state setup. The overlap of the normalized two-photon part of the wavefunction $\kt{\Psi_{2ph}}$ with the two-photon Laughlin state (\ref{eq:psiLaughlin}) is given by $F^2=|\br{\Psi_{2ph}}\Psi_{2L}\rangle|^2$. By varying the atomic density up to 1 $\mu$m$^{-3}$, we see that as the collective light-matter coupling is increased, the state of two photons in the system approaches the Laughlin state with high probability. Coupling photons instead into the $l=N-1$ mode of the lowest Landau level would lead to creation of the $N$-boson Laughlin state with total angular momentum $L=N(N-1)$, due to the vanishing blockade in that state. In the \supp{} we generalize the description of photonic Laughlin states to manifolds with $s>1$.

In conclusion, we have demonstrated that Rydberg polaritons in multimode optical cavities provide a powerful platform for quantum simulation of strongly interacting photonic systems. The dark polaritons are well-described as particles in two dimensions and their Hamiltonian can be controlled through the design of the optical cavity. In particular, degenerate manifolds of Hermite-Gauss modes naturally lead to a harmonic oscillator Hamiltonian, while degenerate Laguerre-Gauss modes in helical cavities naturally give rise to an effective gauge field. The interactions between dark polaritons can be strong enough to induce single-photon nonlinearities, allowing simulation of interacting quantum systems beyond the mean-field regime. In particular, we have shown that this platform allows studies of emergent crystallinity and fractional quantum Hall states. Our work motivates experiments on this system, as well as future theoretical studies of the system. In particular, this system gives access to many-body physics in a driven, dissipative regime, and promises to extend our understanding beyond conventional equilibrium and near-equilibrium studies.

\begin{acknowledgments}
A. T. S. and J. S. acknowledge support from DARPA and AFOSR. H. P. B. acknowledges the support of the Center for Integrated Quantum Science and Technology (IQST), the Deutsche Forschungsgemeinschaft (DFG) within the SFB/TRR 21, and the European Union H2020 FET Proactive project RySQ (grant N. 640378).
\end{acknowledgments}

\vspace{5mm}
\appendix
\section{Theory of Degenerate Cavities}
The frequencies of cavity modes, within the paraxial approximation, are given by~\cite{sieg1986lasers,habr07twist}
\begin{equation}
\omega_{mnp} =\frac{c}{L_r}\left[(m+1/2)\chi_1+(n+1/2)\chi_2 + 2\pi p\right]
\label{eq:resonantfreqsM}
\end{equation}
where $m$ and $n$ are positive integer transverse mode indices, $p$ is the integer longitudinal mode index, $c$ is the speed of light, and $L_r$ is the round-trip cavity length. The Gouy phases $\chi_{1,2}$ are given by the four eigenvalues $e^{\pm i\chi_{1,2}}$ of the round-trip ray matrix.  

Degeneracies occur when the Guoy phases are tuned to rational fractions of $2\pi$. To obtain a two-dimensional harmonic oscillator Hamiltonian, one uses $\chi_1=\chi_2 = 2\pi q/s + \epsilon$. The mode frequencies (\ref{eq:resonantfreqsM}) within a nearly degenerate manifold are then given by
\begin{eqnarray}
	\omega_{mn} &=& \omega_{\bot}(m+n+1) + \mathrm{const}
	\label{eq:harmonicspectrumM}
\end{eqnarray}
with $\omega_{\bot}=\epsilon c/L_r$, and the constraint that $m+n$ mod $s$ is constant within the manifold. 

For a helical cavity, the indices $m$ and $n$ in (\ref{eq:resonantfreqsM}) count units of positive and negative orbital angular momentum in the transverse plane. 
Landau levels are realized when one of the Gouy phases, say $\chi_1$, is tuned to a rational fraction of $2\pi$. The case $\chi_1\rightarrow 0$ results in conventional Landau levels containing a state at each value of the orbital angular momentum. In general, for $\chi_1=2\pi q/s$, each degenerate manifold includes modes with orbital angular momenta in steps of $s$. Tuning slightly away from degeneracy, so that $\chi_1=2\pi q/s+\epsilon$, induces a harmonic potential in addition to the effective magnetic field. 

\section{Numerical Simulation}
To simulate the two-photon crystal and Laughlin states, we obtain the steady-state wavefunction of the system in the limit of a weak probing beam. The system is modeled using 30 atoms positioned randomly within the atomic cloud, which is sufficient to approximate the continuum limit. The coupling strength is scaled up to the appropriate value to simulate a given atomic density. 


We carry out the calculation using perturbation theory to second order in the weak probe field to capture one- and two-photon effects. The probe laser acts through a Hamiltonian term
\begin{equation}
	V = \hbar R \int d\vecrho\:\left[v_L(\vecrho) e^{-i\delta_L t}\psiph^\dag(\vecrho) + \hc\right]
\label{eq:Vdrive}
\end{equation}  
where $R$ is the driving strength, $v_L(\vecrho)$ is the mode profile of the probe laser in the transverse plane of the cavity waist, $\delta_L$ is the detuning of the probe laser from EIT resonance, and $\psiph$ is the photon field operator, defined in greater detail below. We work in the limit of weak driving, $R\rightarrow 0$. In this limit, the total probability of an excitation in the cavity becomes vanishingly small, making the average time between quantum jumps arbitrarily large. The steady state of the system then approaches a pure state~\cite{brech1999cavity}.

We obtain the steady state wavefunction $\kt{\psiexact}$ by perturbation theory in the weak drive $V$,
\begin{equation}
\kt{\psiexact} \approx \left(\mathds{1} +  \Gac \V +  \Gac \V \Gac \V\right)\kt{\psizero}
\label{eq:psipert}
\end{equation}
with 
\begin{equation}
\Gac(\delta_L) = \frac{1}{\hbar\delta_L \Nexc - H}\Q,
\label{eq:green}
\end{equation}
where $\kt{\psizero}$ is the ground state of the atom-cavity system, with no photons in the cavity and all atoms in the ground state, and $\Q=\mathds{1}-\kt{\psizero}\br{\psizero}$ projects onto the space orthogonal to $\kt{\psizero}$. Here $H$ is the atom-cavity system Hamiltonian not including $V$, and $\Nexc$ counts the number of excitation quanta in the system,
\begin{equation}	
	\Nexc = \sum_m \cavAn_m^\dag \cavAn_m + \int d\vecr\left(\phie^\dag\phie + \phir^\dag\phir\right)
\label{eq:Nexc}
\end{equation}

The states resulting from (\ref{eq:psipert}) have up to two excitations. Note that the apparent pattern in (\ref{eq:psipert}) is modified beyond second order by the non-zero second-order shift of the eigenvalues in a discrete system. 


\section{Laughlin States in Generalized Landau Levels}\label{sec:genlandau}
The analog of the lowest Landau level (LLL) consists of the manifold of modes with angular momentum quantum numbers 0, $s$, $2s$, etc. for some positive integer $s$. The conventional LLL corresponds to the case $s=1$. 

We first consider eigenstates of the low-frequency effective Hamiltonian for two dark polaritons with total angular momentum $2s$. The wavefunction is of the form
\begin{equation}
\kt{\Psi} =\left[a\frac{1}{\sqrt{2}}\left(A_{sD}^d\right)^2 + b A_{0D}^d A_{2s,D}^d\right]\kt{0}.
\label{eq:psitrial}
\end{equation}
The eigenstate with zero interaction energy has 
\begin{equation}
a/b = -s!\sqrt{2/(2s)!}
\label{eq:abcoef}
\end{equation}
In position space, the wavefunction becomes
\begin{equation}
\qprod{\vecrho_1,\vecrho_2}{\Psi}\propto\left(z_1^s-z_2^s\right)^2 e^{-(|z_1|^2+|z_2|^2)/w_0^2}
\label{eq:laughlin}
\end{equation}
up to normalization, where $\vecrho_j=(x_j,y_j)$ is the position of the $j$-th dark polariton and $z_j = x_j+iy_j$. This state generalizes the two-boson $\nu=1/2$ Laughlin state. To generate this state experimentally, one can couple photons into the cavity mode with angular momentum $s$; interactions then populate the modes with angular momentum $0$ and $2s$. Two-polariton states with total angular momentum $ns$ for even $n$ can be realized by coupling into the mode with angular momentum $ns/2$, resulting in the analog of the $\nu=1/n$ two-boson Laughlin state $\propto (z_1^s-z_2^s)^n$. 

For $N$ dark polaritons with total angular momentum $L=N(N-1)ns/2$, the eigenstate in position space is
\begin{equation}
\qprod{\vecrho_1,\vecrho_2,...,\vecrho_N}{\Psi}\propto\prod_{i<j}\left(z_i^s-z_j^s\right)^n \prod_k e^{-|z_k|^2/w_0^2}
\label{eq:laughlinmany}
\end{equation}
up to normalization. To realize this state one must couple into the cavity mode with angular momentum $L/N = (N-1)ns/2$.

\section{Cavity Photon Field Operator}
While a photon in an optical cavity moves in a three-dimensional space, the degrees of freedom become two dimensional after restricting to a nearly-degenerate manifold $M$ of the types described in the main text. Two dimensional behavior arises because the longitudinal mode index $p$ becomes a function of the transverse mode indices $m$ and $n$ within the manifold. The photon then has transverse degrees of freedom but no longitudinal degree of freedom.

To define the photon field operator, we first define precisely the transverse mode functions $v_{mn}(x,y)$. In general, the cavity modes are described by orthonormal electric field amplitude functions $u_{mnp}(x,y,z)$. Near a waist, the mode functions factorize into a transverse part and a longitudinal part,
\begin{equation}
	u_{mnp}(x,y,z)\rightarrow v_{mnp}(x,y) w_{mnp}(z)
	\label{eq:factoru}
\end{equation}
which remains valid within the Rayleigh range of the waist. Within a given manifold $M$, as $p$ depends on $m$ and $n$, we write the transverse part simply as $v_{mn}(x,y)$. The functions $v_{mn}(x,y)$ are chosen to be orthonormal,
\begin{equation}
	\int d\vecrho\:d\vecrho' v_{mn}^*(\vecrho)v_{m'n'}(\vecrho')=\delta_{mm'}\delta_{nn'}
	\label{eq:vorth}
\end{equation}
The two-dimensional photon field operator restricted to the manifold $M$ is then defined as 
\begin{equation}
	\psiph(\vecrho) = \sum_{(m,n)\in M}v_{mn}(\vecrho)\cavAn_{mn}
	\label{eq:psiph}
\end{equation}
where $\cavAn_{mn}$ is the annihilation operator for a photon in the $(m,n)$ mode in manifold $M$. The ladder operators have the usual commutation rule $\com{\cavAn_{mn}}{\cavAn_{m'n'}^\dag}=\delta_{mm'}\delta_{nn'}$. 
The commutator of the photon field operators is then
\begin{equation}
\com{\psiph(\vecrho)}{\psiph^\dagger(\vecrho')}=\sum_{(m,n)\in M}v_{mn}(\vecrho)v_{mn}^*(\vecrho')
\label{eq:photoncom}
\end{equation}
which acts as a delta function on the space of functions spanned by $\{v_{mn}\}_{(m,n)\in M}$.

The field operator (\ref{eq:psiph}) allows us to write the second-quantized Hamiltonian for photons in a given manifold,
\begin{eqnarray}
\Hph &=& \sum_{(m,n)\in M}\hbar \omega_{mn} \cavAn_{mn}^\dag\cavAn_{mn}\\
	&=& \int d\vecrho\:\psiph^\dag\,\hph\,\psiph\label{eq:Hph}
\end{eqnarray}
using (\ref{eq:vorth}) and (\ref{eq:psiph}), and defining the first-quantized photon Hamiltonian using $\hph v_{mn} = \hbar\omega_{mn}v_{mn}$.

\section{Two-Dimensional Atomic Excitation Operators}
In this section we provide a precise definition of the two-dimensional collective operators for the atomic excitations. 
To start, we consider the field operator $\Psi_{\alpha}(\vecr)$ for an atom at position $\vecr$ in internal state $\alpha$. As the atom-photon interactions do not create or destroy atoms, but simply change their internal state, we define the local excitation operators
\begin{equation}
	\phi_{\alpha}^\dag(\vecr) = \Psi_\alpha^\dag(\vecr) \Psi_g(\vecr) /\sqrt{n_g(\vecr)}
	\label{eq:phiatom}
\end{equation}
that promote an atom at position $\vecr$ from the ground state to the state $\alpha$. Here $n_g(\vecr)$ is the density of ground state atoms at position $\vecr$. We work in the limit of a  large  ground state density and a small excited state density, so that $n_g$, $\Psi_g$, and $\Psi_g^\dag$ become classical numbers and $\Psi_g, \Psi_g^\dag \rightarrow \sqrt{n_g}$. In this limit, the excitation operators $\phi_\alpha$ obey the commutation rule
\begin{equation}
	\com{\phi_{\alpha}(\vecr)}{\phi_{\alpha'}^\dag(\vecr')} = \delta_{\alpha\alpha'}\delta(\vecr-\vecr')
	\label{eq:phicom}
\end{equation}

To define collective excitation operators, we look at the atom-photon electric dipole interaction,
\begin{equation}
	H_{ed} = \frac{\hbar g_0}{2}\int d\vecr\:\sum_{(m,n)\in M} \sqrt{n_g(z)}\phie^\dag(\vecr)u_{mn}(\vecr)\cavAn_{mn} + \hc
	\label{eq:Hed}
\end{equation}
where $\hbar g_0/2=d_e\sqrt{\frac{\hbar\omega_e}{2\epsilon_0}}$, with $\omega_e$ the resonant frequency of the ground to excited state transition, $d_e$ the dipole matrix element, and $\epsilon_0$ the electric constant.  The atomic density is assumed to be uniform in the transverse dimensions so that $n_g=n_g(z)$. Using the factorization (\ref{eq:factoru}), $H_{ed}$ can be written in terms of the two-dimensional photon field operators (\ref{eq:psiph}) as
\begin{equation}
	H_{ed} = \frac{\hbar g_0}{2}\int d\vecrho\ dz\:\sqrt{n_g(z)}\phie^\dag(\vecrho,z)w(z)\psiph(\vecrho)+\hc
	\label{eq:Hedph}
\end{equation}
where we have used the condition that the atomic sample is thin compared to the Rayleigh range so that the factorization (\ref{eq:factoru}) applies over the relevant range of integration in $z$. Additionally, we have introduced the assumption that the longitudinal part of the cavity mode is independent of the mode index, so that $w_{mn}(z)\equiv w(z)$. This condition holds in a near-planar two-mirror cavity and in travelling-wave cavities with any value of $s$, but not in two-mirror cavities with degeneracies having $s>1$, such as confocal cavities. 

After introducing the two-dimensional field operator for atomic excitations,
\begin{equation}
	\psie^\dag (\vecrho)= \int dz\: \sqrt{\frac{n_g(z)}{\bar{n}}}w(z)\phie^\dag(\vecrho,z)
	\label{eq:psiatom}
\end{equation}
the electric dipole Hamiltonian (\ref{eq:Hedph}) further simplifies to
\begin{equation}
	H_{ed} = \frac{\hbar g_0}{2}\sqrt{\bar{n}}\int d\vecrho\: \psie^\dag(\vecrho)\psiph(\vecrho) + \hc
	\label{eq:Hedpsi}
\end{equation}
The requirement that $\psie$ satisfies the commutation rule
\begin{equation}
	\com{\psie(\vecrho)}{\psie^\dag(\vecrho')}=\delta(\vecrho-\vecrho')
	\label{eq:psipcom}
\end{equation}
fixes the value of the constant $\bar{n}$ at
\begin{equation}
	\bar{n} = \int dz\: n_g(z)|w(z)|^2
	\label{eq:nbar}
\end{equation}
For a sample with thickness $d$ in the $z$ direction, $\bar{n}\sim n_g d /L$. Equation (\ref{eq:Hedpsi}) shows that the collective Rabi frequency is given by $G= g_0\sqrt{\bar{n}}$.

While cavity photons in a given manifold only excite atoms into collective states with a particular longitudinal mode structure, the atomic gas supports arbitrary collective excitations. Additional atomic field operators account for these modes. Defining $\bar{w}_0\equiv w\sqrt{n_g/\bar{n}}$, consider a complete orthonormal set of functions $\{\bar{w}_j\}_{j=0}^\infty$ and let
\begin{equation}
	\psi_{ej}^\dag(\vecrho)=\int dz\:\bar{w}_j(z)\phie^\dag(\vecrho,z)
	\label{eq:psipj} 
\end{equation}
so that $\psi_{e0} = \psie$ from (\ref{eq:psiatom}). The $\psi_{ej}$ operators allow expression of arbitrary three-dimensional operators. For example, the density of atoms in the intermediate state is expressed as
\begin{equation}
	\int d\vecr\:\phie^\dag \phie = \sum_{j=0}^{\infty}\int d\vecrho\: \psi_{ej}^\dag\psi_{ej}
\end{equation}
where we have use the completeness relation
\begin{equation}
	\sum_{j=0}^{\infty} \bar{w}^*_j(z) \bar{w}_j(z') = \delta(z-z')
\end{equation}

The EIT control field couples the collective excitation generated by absorbing a cavity photon to a specific collective excitation of the Rydberg state. We therefore define two-dimensional collective operators $\psi_r(\vecrho)$ for Rydberg excitations in a manner analogous to that of the intermediate state. We identify the correct operators by looking at the Hamiltonian term for the coupling from the intermediate state to the Rydberg state in the rotating wave approximation,
\begin{equation}
	H_c =\hbar\frac{\Omega}{2} \int d\vecr\: e^{i\veckc\cdot \vecr}\phi_r^\dag \phie + \hc
	\label{eq:Hc}
\end{equation}
where $\veckc$ is the wavevector of the control beam. We define the two-dimensional field operators for the Rydberg level as
\begin{equation}
	\psi_{rj}^\dag(\vecrho)=\int dz\:\bar{w}_j(z)e^{i\veckc\cdot\vecr}\phir^\dag(\vecrho,z)
	\label{eq:psirj}
\end{equation}
The control field coupling (\ref{eq:Hc}) becomes
\begin{equation}
	H_c = \hbar\frac{\Omega}{2} \sum_{j=0}^\infty\int d\vecrho\:\psi_{rj}^\dag\psi_{ej} + \hc
	\label{eq:Hctwo}
\end{equation}

The two-dimensional field operators $\psi_{ej}$ and $\psi_{rj}$ satisfy the commutation rule
\begin{equation}
	\com{\psi_{\alpha j}(\vecrho)}{\psi_{\alpha' j'}^\dag(\vecrho')} = \delta_{\alpha\alpha'}\delta_{jj'}\delta(\vecrho-\vecrho')
	\label{eq:psipjcom}
\end{equation}
with $\alpha,\alpha' =$ e, r. 

The Hamiltonian for Rydberg EIT in a nearly-degenerate multimode cavity can then be written in therms of the two-dimensional excitation operators as  
\begin{eqnarray}
	\Htotal &=& \hbar\int d\vecrho
	\colthree{\psiph}{\psie}{\psir}^\dag\!
	\left(\begin{array}{ccc}
		\hphtilde & G/2 & 0\\
		G/2  & \Deltatilde & \Omega/2\\
		0 & \Omega/2 & -i\gamma/2\\
	\end{array}\right)
	\colthree{\psiph}{\psie}{\psir}\nonumber\\
	&&+\hbar\sum_{j=1}^\infty\int d\vecrho\:\coltwo{\psi_{ej}}{\psi_{rj}}^\dag
	\left(\begin{array}{cc}
		\Deltatilde & \Omega/2\\
		\Omega/2 & -i\gamma/2\\
	\end{array}\right)\coltwo{\psi_{ej}}{\psi_{rj}}\nonumber\\
	&&+\Hint
	\label{eq:fullH}
\end{eqnarray}
with $\hphtilde=\hph/\hbar-i\kappa/2$. The second line in (\ref{eq:fullH}) accounts for matter modes with longitudinal structures that do not couple to the cavity. These modes contribute to the screening of the polariton-polariton interaction.

\section{Diagonalization of the Single Polariton Hamiltonian}
The single-particle eigenstates of the system Hamiltonian (\ref{eq:fullH}) can be found by working in a basis defined by the cavity modes. To express atomic excitations that do not couple to any cavity mode in the chosen manifold $M$, we extend the basis defined by $M$ to a complete orthonormal set of modes. The collective operators for excitations coupled to an arbitrary mode $m$ are then defined as
\begin{equation}
	D_m = \int d\vecr\: \phie \bar{u}_m^* 
	\label{eq:Dm}
\end{equation}
for the intermediate state, and
\begin{equation}
	F_m = \int d\vecr\: e^{-i\veckc\cdot \vecr}\phir \bar{u}_m^*
	\label{eq:Fm}
\end{equation}
for the Rydberg state. Here $\bar{u}_m(\vecr) \equiv u_m(\vecr)\sqrt{n_g/\bar{n}}$ for $m\in M$. For $m\notin M$, the $\bar{u}_m$ are a basis for the orthogonal complement of the cavity modes in the space of normalizable functions, so that $\{\bar{u}_m\}$ taken over all $m$ form a complete, orthonormal basis. The collective mode operators obey $\com{D_m}{D_n^{\dagger}}=\com{F_m}{F_n^{\dagger}}=\delta_{mn}$.

The fundamental form of the system Hamiltonian is given by
\begin{eqnarray}
	\Htotal &=& \hbar\sum_{m\in M} \left\{\delta_m \cavAn_m^\dagger\cavAn_m+
	\int d\vecr\: \frac{g_0}{2}\sqrt{n_g}\left[\phie^\dagger u_m\cavAn_m + \hc
	\right]
	\right\}\nonumber\\
	& & +\hbar\int d\vecr\:\coltwo{\phie}{\phir}^\dagger
	\left(
	\begin{array}{cc}
	\Delta & e^{-i\veckc\cdot \vecr}\Omega/2\\
	e^{i\veckc\cdot \vecr}\Omega/2 & 0 \\
	\end{array}
	\right)\coltwo{\phie}{\phir}\nonumber\\
	& &+\Hint
	\label{eq:Hfundamental}
\end{eqnarray}
where $\delta_m$ is the detuning of mode $m$ from EIT resonance. 
Using the collective operators (\ref{eq:Dm}) and (\ref{eq:Fm}), the non-interacting part of the system Hamiltonian $\Hnon\equiv \Htotal-\Hint$ can be written as 
\begin{eqnarray}
	\Hnon &=& \hbar\sum_{m\in M}\colthree{\cavAn_m}{D_m}{F_m}^\dagger
	\left(
	\begin{array}{ccc}
	\tilde{\delta}_m & G/2 & 0\\
	G/2 & \tilde{\Delta} & \Omega/2\\
	0 & \Omega/2 & -i\gamma/2\\
	\end{array}
	\right)
	\colthree{\cavAn_m}{D_m}{F_m}\nonumber\\
	&&+\hbar\sum_{m\notin M}\coltwo{D_m}{F_m}^\dagger
	\left(
	\begin{array}{cc}
	\tilde{\Delta}& \Omega/2\\
	\Omega/2 & -i\gamma/2\\
	\end{array}
	\right)
	\coltwo{D_m}{F_m}\label{eq:Hcollective}
\end{eqnarray}

The single-excitation eigenstates of $\Htotal$ in the photonic modes $m\in M$ are familiar EIT eigenstates~\cite{flei00dark}, obtained by diagonalizing the 3x3 matrix in (\ref{eq:Hcollective}). The two bright states have complex eigen-frequencies $\tilde{\delta}_{mB\pm}\approx\left(\tilde{\Delta}\pm\sqrt{\tilde{\Delta}^2+G^2+\Omega^2}\right)/2$, to zeroth order in $\delta_m$ and $\gamma$. The dark state eigen-frequencies $\tilde{\delta}_{mD}=\delta_{mD} - i\gamma_{mD}/2$ are given by
\begin{eqnarray}
\tilde{\delta}_{mD} &\approx& \frac{\Omega^2}{G^2+\Omega^2}\left(\delta_m-i\frac{\kappa_m}{2}\right)
	+\frac{G^2}{G^2+\Omega^2}\left(-i\frac{\gamma}{2}\right) \nonumber \\
	&&+\frac{4G^2\Omega^2(\Delta-i\frac{\Gamma}{2})}{(G^2+\Omega^2)^3}
	\left(\delta_m+i\frac{\gamma-\kappa_m}{2}\right)^2
\label{eqn:darkeig}
\end{eqnarray}
to second order in the small quantities $\delta_m$, $\kappa_m$, and $\gamma$. The leading order probability for a dark state polariton to be a photon is $\cos^2\theta=\Omega^2/(G^2+\Omega^2)$, where $\theta$ is the dark state rotation angle. The first term in (\ref{eqn:darkeig}) therefore gives the probability for the dark polariton to be a photon, times the complex detuning of the photon, while the second term gives the probability for the dark state polariton to be a Rydberg excitation, times the loss in the Rydberg state. The second line in (\ref{eqn:darkeig}) gives the second order eigenvalue, reflecting modifications to the dark state wavefunction due to detuning and losses. In particular, a non-zero detuning $\delta_m$ from EIT resonance leads to population of the atomic intermediate state and to a loss rate proportional to $\Gamma \delta_m^2$.

\subsubsection{Dark Polaritons}
We diagonalize the 3x3 matrix in (\ref{eq:Hcollective}) as $\mat{H}_m = \mat{Q}_m \mat{E}_m \mat{Q}_m^{-1}$, where 
$\mat{E}_m=\mathrm{diag}(\tilde{\delta}_{mD},\tilde{\delta}_{m+},\tilde{\delta}_{m-})$. 
The columns of $\mat{Q}_m$ are right eigenvectors of $\mat{H}_m$ while the rows of $\mat{Q}_m^{-1}$ are left eigenvectors of $\mat{H}_m$. The annihilation operators for dark and bright polaritons in mode $m$ are then
$(A_{mD},A_{m+},A_{m-})'\equiv\mat{A}_m=\mat{Q}_m^{-1}\mat{J}_m$, with $\mat{J}_m=(c_m,D_m,F_m)'$. 
The dual creation operators are 
$(A_{mD}^d,A_{m+}^d,A_{m-}^d)\equiv\mat{A}_m^d=\mat{J}_m^\dagger\mat{Q}_m$.
The polariton operators obey the commutation rules $\com{A_{m\mu}}{A_{n\nu}^d}=\delta_{mn}\delta_{\mu\nu}$ and $\com{A_{m\mu}}{A_{n\nu}}=\com{A_{m\mu}^d}{A_{n\nu}^d}=0$.

Field operators for the polariton excitations are then defined as $\psi_{\mu} = \sum_{m\in M}v_m A_{m\mu}$ and $\psi_{\mu}^d=\sum_{m\in M}v_m^* A_{mD}^d$, for $\mu\in\{D,+,-\}$. The commutator for the field operators is $\com{\psi_{\mu}(\vecrho)}{\psi_{\nu}^d(\vecrho')}=\delta_{\mu\nu}\sum_{m\in M}v_m(\vecrho)v_m^*(\vecrho')$; the sum acts as a delta function on the space of functions spanned by $\{v_m\}$.

%
%
The term in the Hamiltonian (\ref{eq:Hcollective}) describing the dark polariton excitations then becomes 
\begin{eqnarray}
	\tilde{H}_D &\equiv& \sum_{m\in M}A_{mD}^d \tilde{\delta}_{mD} A_{mD} \\
	&=&\int d\vecrho\: \psi_D^d \HtildediffD(\vecrho) \psi_D\label{eq:HtildeD}
\end{eqnarray}
where $\HtildediffD(\vecrho)$ is a differential operator on functions of the transverse coordinates $\vecrho=(x,y)$ such that 
\begin{equation}
\HtildediffD v_m=\hbar \tilde{\delta}_{mD}v_m
\label{eq:mathcalHD}
\end{equation}

From (\ref{eqn:darkeig}) we see that $\HtildediffD$ is given to leading order by $\HdiffD\equiv \hph\:\cos^2\theta $, with $\hph$ the photon Hamiltonian introduced in (\ref{eq:Hph}). 
For the harmonic oscillator-like and Landau level-like manifolds of cavity modes,
\begin{equation}
\HdiffD(\vecrho) = \frac{1}{2 m_D}\left(-i\hbar\partial_{\vecrho}-\vecA\right)^2 + \frac{1}{2}m_D\omega_{\bot D}^2\rho^2,
\label{eq:HdiffD}
\end{equation}
where the effective gauge field $\vecA\neq 0$ for the Landau level-like case. Note that the polariton mass is increased relative to the photon mass by the factor $1/\cos^2\theta$ and the oscillation frequency is decreased by $\cos^2\theta$ so each term in (\ref{eq:HdiffD}) equals the corresponding term in the photon Hamiltonian, scaled by $\cos^2\theta$.

\section{Two-Polariton Problem}\label{sec:interact}
The effective interaction between dark polaritons follows from the solution of the two-body problem
\begin{equation}
	\hbar\omega\kt{\Psi}=(\Hnon+\Hint)\kt{\Psi}
	\label{eq:schrodinger}
\end{equation}
where $\kt{\Psi}$ is an arbitrary two-excitation state.

The problem can be re-written as 
\begin{equation}
\kt{\Psi}=G_0(\omega)\Hint\kt{\Psi}
\label{eq:Gschrodinger}
\end{equation}
where
\begin{equation}
G_0(\omega)=\left[\hbar\omega-\Hnon\right]^{-1}
\label{eq:Gzero}
\end{equation}
The inverse exists for most values of $\omega$ because this is a discrete system. Equation (\ref{eq:Hcollective}) shows that $H_0$ 
is block diagonal in pairs of modes $(m,n)$. We find the Green function (\ref{eq:Gzero})
by inverting within each block. 

As the interaction term $\Hint$ projects $\kt{\Psi}$ onto the subspace containing two Rydberg excitations, projecting (\ref{eq:Gschrodinger}) onto the two-Rydberg subspace gives a Schr\"odinger equation for Rydberg-Rydberg part of the wavefunction alone,
\begin{equation}
	\varphi_{rr}= G_{rr} \Hint\varphi_{rr}
	\label{eq:phiss}
\end{equation}
where $\varphi_{rr}$ is the Rydberg-Rydberg component of $\kt{\Psi}$ and $G_{rr}$ is the component of $G_0$ that takes two Rydberg excitations to two Rydberg excitations. 
In the mode basis, $G_{rr}$ is diagonal in pairs of modes. Its elements consist of three types, corresponding to three cases for the pairs of modes: neither in $M$, one in $M$ and one not in $M$, and both in $M$. We decompose $G_{rr}$ into these three types using projection operators,
\begin{equation}
 G_{rr}= \bar{\chi}P + \sum_{m\in M}\chi_1(\delta_m)P_m  +\sum_{m,n\in M} \chi_0(\delta_m,\delta_n)P_{mn}
\end{equation}
where $P_{mn}$ projects onto the pair of modes $m$ and $n$. In the position basis,
\begin{equation}
	P_{mn}(x,y;x',y') = \bar{u}_m(x) \bar{u}_m^*(x') \bar{u}_n(y) \bar{u}_n^*(y')
	\label{eq:Pmn}
\end{equation}
in terms of which the two other projectors are
\begin{eqnarray}
P_m&=& \sum_{n\notin M}(P_{mn}+P_{nm})\\
P&=&\sum_{m,n\notin M}P_{mn}
\end{eqnarray}
The projectors satisfy the completeness relation:
\begin{equation}
P+\sum_{m\in M} P_m + \sum_{m,n\in M}P_{mn} = \mathds{1},
\end{equation}
where $\mathds{1}$ is the identity operator on the space of two-Rydberg wavefunctions. 

We obtain the effective potential by subtracting $\bar{\chi}\mathds{1}$ from $G_{rr}$,
\begin{eqnarray}
\Delta G &=& G_{rr} - \bar{\chi}\mathds{1}\\
&=& \sum_{m\in M}\left[\chi_1(\delta_m)-\bar{\chi}\right] P_m\nonumber\\
&&+\sum_{m,n\in M}\left[\chi_0(\delta_m,\delta_n)-\bar{\chi}\right]P_{mn}
\label{eq:deltag}
\end{eqnarray}
The two-Rydberg component (\ref{eq:phiss}) of the Schr\"odinger equation can then be re-written
\begin{equation}
\tilde{\varphi} = \Delta G \:\Ueff \tilde{\varphi}.
\label{eq:effschro}
\end{equation}
with,
\begin{equation}
\Ueff = \frac{V}{1-\bar{\chi}V}\quad \mathrm{and} \quad
\varphi_{rr} =\frac{\tilde{\varphi}}{1-\bar{\chi}V}
\label{eq:tildephi}
\end{equation} 

Equation (\ref{eq:effschro}) implies $P\tilde{\varphi}=0$. Therefore we expand,
\begin{equation}
\tilde{\varphi}(\vecx,\vecy)=\frac{1}{2}\sum_{m\in M}\left[\bar{u}_m(\vecx)\theta_m(\vecy)+\theta_m(\vecx)\bar{u}_m(\vecy)\right],
\label{eq:phiexpand}
\end{equation}
with $\theta_m$ as arbitrary functions,
\begin{equation}
\theta_m(\vecr)=\sum_n a_{mn}\bar{u}_n(\vecr)
\label{eq:theta}
\end{equation}

The Schr\"odinger equation (\ref{eq:effschro}) can be rewritten as an algebraic equation for the expansion coefficients $a_{mn}$, giving
\begin{equation}
a_{kl}=\left[\chi_1(\delta_k)-\bar{\chi}\right]\sum_{m\in M; n} a_{mn}(A_{kl}^{mn}+A_{kl}^{nm})\label{eq:MN}
\end{equation}
for $k\in M, l\notin M$ and 
\begin{equation}
a_{kl}+a_{lk} = \left[\chi_0(\delta_k,\delta_l)-\bar{\chi}\right]\sum_{m\in M;n}a_{mn}(A_{kl}^{mn}+A_{kl}^{nm})
\label{eq:MM}
\end{equation}
for $k,l\in M$.
The matrix elements of the effective potential are defined as,
\begin{eqnarray}
A_{kl}^{mn}&=&\int d\vecx d\vecy\:\bar{u}_k^*(\vecx)\bar{u}_l^*(\vecy)\Ueff(\vecx-\vecy)\bar{u}_m(\vecx)\bar{u}_n(\vecy)\nonumber\\
&=& A_{lk}^{nm}\label{eq:Aklmn}
\end{eqnarray}

\subsubsection{Low-Energy Limit}
When the interaction is sufficiently weak, or when there are sufficiently many modes so that the excitations can be spatially separate, the Hamiltonian has small eigenvalues $\omega$. In the $\omega\rightarrow 0$ limit, $\chi_0(\delta_m,\delta_n)$ becomes large for small $\delta_m,\delta_n$ and dominates over the other terms $\chi_1(\delta_k)$ and $\bar{\chi}$. Equations (\ref{eq:MN}) and (\ref{eq:MM}) give
\begin{equation}
	a_{kl}\rightarrow 0
	\label{eq:MNlow}
\end{equation}
for $k\in M, l\notin M$ and 
\begin{equation}
	a_{kl}+a_{lk} \rightarrow \chi_0(\delta_k,\delta_l)\sum_{m,n\in M}(a_{mn}+a_{nm})A_{kl}^{mn}
	\label{eq:MMlow}
\end{equation}
For $k,l\in M$.
Also, $\tilde\varphi$ becomes
\begin{equation}
\tilde\varphi(x,y)\rightarrow\frac{1}{2}\sum_{m,n\in M}(a_{mn}+a_{nm})\bar{u}_m(x) \bar{u}_n(y)
\label{eq:philow}
\end{equation}
So only the sum $a_{mn}+a_{nm}$ has significance. Define
\begin{equation}
b_{mn}= \frac{1}{2}(a_{mn}+a_{nm}).
\label{eq:b}
\end{equation}
Also note that the asymptotic form of $\chi_0$ is
\begin{equation}
\chi_0(\delta_k,\delta_l) \rightarrow \frac{\alpha}{\omega-\en_k-\en_l}\quad \mathrm{with}\quad \alpha=\frac{G^4}{(G^2+\Omega^2)^2}
\label{eq:chizerolow}
\end{equation}
and $\en_k$ the eigenvalue of the single-polariton Hamiltonian for a dark polariton in mode $k$.

The low-energy Schr\"odinger equation (\ref{eq:MMlow}) becomes
\begin{equation}
\omega\: b_{kl}=(\en_k+\en_l) b_{kl} + \alpha\sum_{m,n\in M}A_{kl}^{mn} b_{mn}
\label{eq:blow}
\end{equation}

The low-energy effective Schr\"odinger equation (\ref{eq:blow}) describe excitations in the cavity modes interacting via the effective potential $\Ueff$. 

Finally we show that the full eigenstate $\psi$ simply describes a pair of dark polaritons in a superposition of cavity modes, via 
\begin{equation}
\psi = G_0 \Ueff \tilde\varphi
\label{eq:psi}
\end{equation}
which requires the use of all rows of the two-Rydberg column of each block of $G_0$. In the limit where the $\en_k$ are smaller than the $\alpha A_{kl}^{mn}$, $\tilde\varphi$ is a function of the restriction of $\Ueff$ to $M$. Since $G_0$ suppresses the non-$M$ components, as well as the non-dark polariton components, the end result of (\ref{eq:psi}) is a superposition of pairs of dark polaritons in the same modes described by $\tilde\varphi$.  However, when the $\en_k$ are larger, $\Ueff$ mixes the modes of $\varphi$. 

\subsubsection{Polariton-Polariton Interaction}
So far we have obtained an effective potential $\Ueff$ in three dimensions. To define the effective potential between dark polaritons we note that the interaction matrix elements $A_{kl}^{mn}$ of Eq. (\ref{eq:Aklmn}) can be written, when all the indices refer to modes in $M$, as 
\begin{equation}
A_{kl}^{mn}=\int d\vecrho d\vecrho'\:v_k^*(\vecrho) v_l^*(\vecrho')\Uefftwo(\vecrho-\vecrho')v_m(\vecrho)v_n(\vecrho')
\label{eq:Atwo}
\end{equation}
with
\begin{equation}
\Uefftwo(\vecrho) =\int dz dz' |\bar{w}(z) \bar{w}(z')|^2 \:\Ueff(\vecrho,z-z')
\end{equation}
In the limit of a sample that is thin compared to the blockade radius,
\begin{equation}
\Uefftwo(\vecrho)\rightarrow \Ueff(\vecrho,0)
\end{equation}
Finally, the interaction between dark polaritons is 
\begin{equation}
U_D(\vecrho) = \alpha \Uefftwo(\vecrho)
\label{eq:UD}
\end{equation}
We can then write the interaction between dark polaritons in second quantized notation,
\begin{eqnarray}
	H_D &=& \int d\vecrho \psi_D^\dag \hdark \psi_D \label{eq:Heff} \\
		&&+\frac{1}{2}\int d\vecrho d\vecrho' \psi_D^\dag(\vecrho) \psi_D^\dag(\vecrho') U_D(\vecrho-\vecrho')\psi_D(\vecrho)\psi_D(\vecrho')
		\nonumber
\end{eqnarray}

 %
%

\end{document}